\documentclass[12pt]{iopart}

\usepackage{iopams}

\expandafter\let\csname equation*\endcsname=\relax
\expandafter\let\csname endequation*\endcsname=\relax
\usepackage{amsmath}

\usepackage{amssymb}

\usepackage{graphicx}

\usepackage{color}
\usepackage[colorlinks]{hyperref}

\begin{document}

\title[Viscocapillary Instability in Cellular Spheroids]{Viscocapillary Instability in Cellular Spheroids}

\author{Matthieu Martin$^{1}$ and Thomas Risler$^{1}$}
\address{$^1$Laboratoire Physico-Chimie Curie, Institut Curie, PSL Research University, Sorbonne Universit\'e, CNRS, 26 rue d'Ulm, 75005 Paris, France}
\ead{thomas.risler@curie.fr}

\begin{abstract}
We describe a viscocapillary instability that can perturb the spherical symmetry of cellular aggregates in culture, also called multicellular spheroids. In the condition where the cells constituting the spheroid get their necessary metabolites from the immediate, outer microenvironment, a permanent cell flow exists within the spheroid from its outer rim where cells divide toward its core where they die. A perturbation of the spherical symmetry induces viscous shear stresses within the tissue that can destabilise the aggregate. The proposed instability is viscocapillary in nature and does not rely on external heterogeneities, such as a pre-existing pattern of blood vessels or the presence of a substrate on which the cells can exert pulling forces. It arises for sufficiently large cell-cell adhesion strengths, cell-renewal rates, and metabolite supplies, as described by our model parameters. Since multicellular spheroids in culture are good model systems of small, avascular tumours, mimicking the metabolite concentration gradients found {\it in vivo}, we can speculate that our description applies to microtumour instabilities in cancer progression.
\end{abstract}

\submitto{\NJP}

\maketitle

\section{Introduction}

Interface instabilities in systems driven far from equilibrium have been extensively studied in solid-state physics. Classical examples are the Saffman-Taylor instability, which occurs when a fluid of lower viscosity displaces a more viscous one in a Hele-Shaw cell~\cite{SaffmanPenetration1958,BensimonViscous1986}, the Mullins-Sekerka instability, which stems from the diffusive transport of the latent heat of solidification in unidirectional solidification~\cite{MullinsStability1964,LangerInstabilities1980}, and the Rayleigh-Taylor instability, corresponding to the fingering of an interface between two immiscible fluids of different densities when the heavier fluid is placed on top of the lighter~\cite{LordRayleighScientific1900,TaylorInstability1950}. 

Instabilities originating from similar coupling terms as those responsible for these classical condensed-matter instabilities have been identified in living systems. In tissues or bacterial colonies, growth and cell divisions may give rise to similar or new out-of-equilibrium phenomena~\cite{RislerHomeostatic2015,KalziqiImmotile2018,WilliamsonStability2018,AlertActive2019}. For example, in the case of bacterial-colony growths, patterns similar to those associated with aggregation phenomena and viscous fingering have been observed~\cite{Ben-JacobGeneric1994,Ben-JacobCooperative2000}. Such a coupling can lead to fractal branching patterns via the process of diffusion-limited aggregation~\cite{SanderFractal1986,FujikawaFractal1989} or other types of branching patterns, depending on the bacterial morphotype~\cite{Ben-JacobCooperative2000}. In tissues, mechanical instabilities have been recognised to play a potential role in different morphological processes and patterns exhibited by growing cell populations~\cite{WangThreedimensional2015}. Examples are the wrinkling patterns of growing, soft surfaces~\cite{LiMechanics2012,WangThreedimensional2015} such as those of leaves and flowers~\cite{DervauxMorphogenesis2008,LiangGrowth2011}, the large-scale looping morphology of the gut~\cite{SavinGrowth2011,BenAmarAnisotropic2013,DurelMechanobiology2020} and the generation of its surface villi~\cite{ShyerVillification2013,DurelMechanobiology2020}, or the formation of cortical convolutions~\cite{TallinenGrowth2016}. Such instabilities may emerge from a buckling phenomenon~\cite{DrasdoBuckling2000,HannezoMechanical2012}, potentially driven by the differential growth of adjacent tissue layers~\cite{KuckenFingerprint2005,LiangGrowth2011,TozluogluPlanar2019,EngPlant2020}, or by other curling and crumpling instabilities due to anisotropic growth~\cite{DervauxMorphogenesis2008}.

Here, we focus on the stability of the spherical growth of cellular aggregates. Such experimental systems are used to study the growth dynamics and cellular structure of microscopic tumours with realistic metabolite concentration gradients~\cite{SutherlandCell1988}. They have also been used as anti-cancer therapy test platforms, mirroring the three dimensional cellular context and therapeutically relevant pathophysiological gradients of {\it in-vivo} tumours~\cite{HirschhaeuserMulticellular2010,Costa3D2016}. Cellular aggregates also permit the study of the effects of different perturbations on the growth or cellular-duplication dynamics, such as a change in the external mechanical constraints~\cite{HelmlingerSolid1997,MontelStress2011,DelarueMechanical2013,KhalifatSoft2016,DolegaCelllike2017}. Most often, an effective surface tension exists between the aggregate and its direct environment, which makes it look like a spheroid, justifying the denomination of multicellular spheroid~\cite{ForgacsViscoelastic1998,LecuitCell2007,GuevorkianAspiration2010,ManningCoaction2010,Gonzalez-RodriguezSoft2012}. The supply of metabolites from the microenvironment is responsible for an inhomogeneous distribution of cell divisions within the spheroid, with an increased cell proliferation at its periphery and an increased cell death in its core~\cite{HirschhaeuserMulticellular2010}. As a consequence, cells flow from the outer rim toward the centre of the spheroid~\cite{DorieMigration1982,DelarueMechanical2013}. Under some circumstances, the spheroid reaches a steady-state size~\cite{JiangMultiscale2005,ByrneDissecting2010,DelarueMechanical2013}.

In the present paper, we investigate theoretically whether this steady-state spherical shape can be unstable without changing the average spheroid size, due to the shear stresses created by differential inward-directed cell flows. Similar shape instabilities have been proposed in the context of the growth of small, spherical tumours, starting with the seminal work of H.P.~Greenspan~\cite{GreenspanGrowth1976}. These instabilities can be described in terms of reaction-diffusion processes, driven by the transport of growth-affecting factors. Some factors such as nutrients, acidity, or chemotherapeutic agents diffuse from external sources to the tumour cells~\cite{FerreiraReactiondiffusion2002,CristiniNonlinear2003,MacklinNonlinear2007}, while others are produced by the tumour cells themselves in a positive or negative feedback loop~\cite{ChaplainSpatiotemporal2001,KhainDynamics2006}. Other descriptions are based on external adhesion cues or elastic heterogeneities. Cell-cell adhesions can participate in maintaining tumour compactness and radially symmetric geometry~\cite{ByrneModelling1996,FrieboesIntegrated2006} but may drive phase separations between different sub-populations of cells~\cite{ChatelainEmergence2011}. Differential growth processes in heterogeneous elastic media or differential pulling forces can also trigger shape instabilities in two or three dimensions~\cite{DrasdoModeling2012,CiarlettaBuckling2013}. Most of the aforementioned instabilities however rely on cell migration, elastic or poroelastic tissue models, or numerical simulations. For a multicellular spheroid grown in a culture medium, however, there is no external medium on which the cells can pull, and only internal stresses can contribute to a potential shape instability. Also, experiments and modelling suggest that, on long timescales, cell-cell rearrangements lead to an effective viscous rheology of the tissue~\cite{ForgacsViscoelastic1998,MarmottantRole2009,RanftFluidization2010,Gonzalez-RodriguezSoft2012,PopovicActive2017,PetridouFluidizationmediated2019}, and an effective surface tension at tissue boundaries has been shown to play an important role in their shaping~\cite{LecuitCell2007,SchotzQuantitative2008,MgharbelMeasuring2009,GuevorkianAspiration2010,ManningCoaction2010}.

\begin{figure}[t]
\begin{center}
\includegraphics[width=0.75\textwidth]{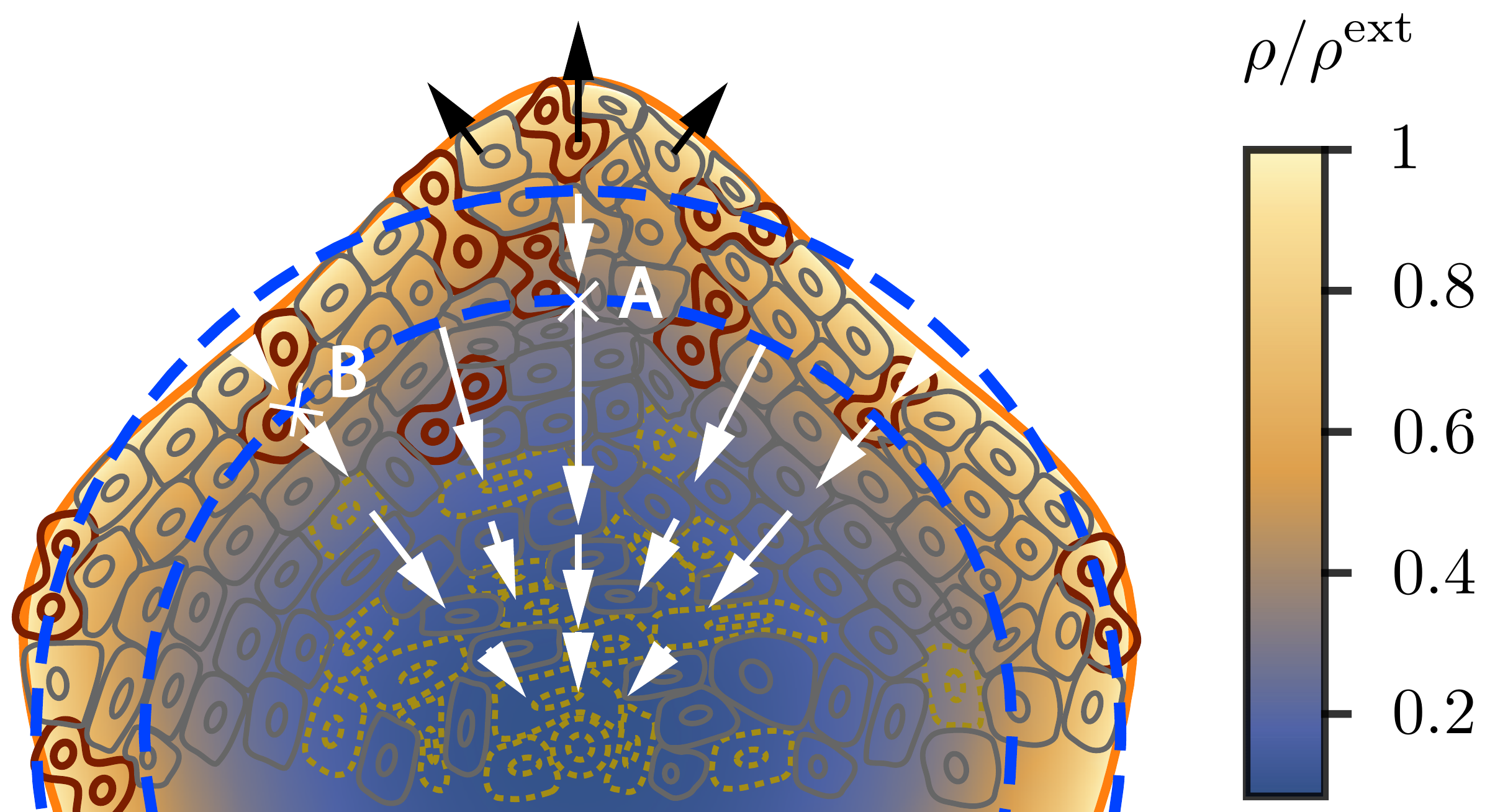}
\end{center}
\caption
{\label{figSchematic} Schematic drawing of the proposed viscocapillary instability in a multicellular spheroid. The original spherical shape is indicated by the external blue-dashed circle. The scaled metabolite concentration is depicted as a scaled colour gradient. Dividing cells (red) are located mostly close to the outer surface. Close to the centre, cells are deprived of metabolites and die (yellow dashed). An inner blue circle indicates cells equidistant from the spheroid centre, and the tissue-flow field is qualitatively depicted by white (resp. black) arrows for inward (resp. outward) cell velocities. At point {\bf A}, cells are pushed inward faster than at point {\bf B}, because more layers of dividing cells push the cells in {\bf A} toward the spheroid core. This creates viscous shear stresses that, in reaction, push the cells located above {\bf A} outwards. Capillary effects due to tissue surface tension at the outer boundary, however, favour stability. Depending on parameter values, the original spherical shape can or not become unstable by this viscocapillary mechanism.}
\end{figure}

In the present work, we rely on a viscous description of cellular tissues on long timescales to establish analytically a new instability of multicellular spheroids that is viscocapillary in nature. This new instability does not require external heterogeneities or the presence of a substrate on which cells could exert pulling forces. Rather, it is powered by viscous shear stresses that build up within the spheroid due to permanent cell renewal, as illustrated in figure~\ref{figSchematic}.
The instability can be triggered by a change of internal properties such as cell-cell adhesion strength, cell-renewal rate, or metabolite supply.

\section{Description of the model}\label{secModelDesc}

We consider a multicellular spheroid embedded in a culture medium with a constant, physiological external pressure $P^{\rm ext}$, and which contains a given concentration $\rho^{\rm ext}$ of a chemical substance necessary for cellular proliferation. This substance can be oxygen, growth factors, glucose or other nutrients, and will be referred to with the generic term of `metabolites' in the following. The tissue within the spheroid is characterised in a continuum theory by a cell-number density and a cell-velocity field $\mathbf{v}$. Analyses of the stress distribution in multicellular spheroids have shown that, at steady state under physiological osmotic conditions, the cell density is essentially homogeneous throughout one aggregate~\cite{MontelIsotropic2012,DelarueStress2014}. We therefore consider an incompressible tissue, for which the cell-number density is constant and the continuity equation reads
\begin{equation}\label{eqContinuity}
\nabla\cdot\mathbf{v} = k_{\rm p}\, .
\end{equation}
Here, $k_{\rm p}$ is the overall cell-production rate, considering cell division and cell death, and $\nabla$ denotes the spatial derivative operator, contracted with the cell-velocity field to give its divergence. Neglecting inertia and in the absence of bulk external forces, force balance reads
\begin{equation}\label{eqForceBalance}
\nabla\cdot \boldsymbol{\sigma}= 0\, ,
\end{equation}
where $\boldsymbol{\sigma}$ denotes the total stress tensor. We further split the stress tensor into a dynamic part and a velocity-independent part. For an isotropic tissue, the latter reads $-P\,\mathbf{1}$, where $P$ is the tissue pressure and $\mathbf{1}$ the unity tensor. The dynamic part $\boldsymbol{\sigma}'$ encodes the rheological properties of the tissue. The timescale of interest here is large compared to those of individual cellular processes, such as cell-cell rearrangements and cell renewal. We can therefore model the tissue as a viscous fluid with effective shear and bulk viscosities $\eta$ and $\zeta$, taking into account the long-term effects of cell production~\cite{ForgacsViscoelastic1998,MarmottantRole2009,RanftFluidization2010}. We obtain
\begin{equation}\label{eqDeviatoricStress}
\boldsymbol{\sigma}'=\eta \left[ \nabla\otimes\mathbf{v}+(\nabla\otimes\mathbf{v})^{\rm T} -\frac{2}{3}(\nabla\cdot\mathbf{v})\,\mathbf{1}\right]+\zeta(\nabla\cdot\mathbf{v})\,\mathbf{1}\, ,
\end{equation}
where $\otimes$ denotes the tensorial product and $(\nabla\otimes\mathbf{v})^{\rm T}$ the transposed tensor of $\nabla\otimes\mathbf{v}$. The remarkable absence of any compression modulus is due to the nonconservation of cell number~\cite{BasanHomeostatic2009,RanftFluidization2010}.

The system of equations is closed by specifying the expression of the cell-production rate $k_{\rm p}$. For simplicity, we assume that $k_{\rm p}$ is independent of the stress $\boldsymbol{\sigma}$ and linearly dependent on the metabolite concentration $\rho$. This leads to
\begin{equation}\label{eqCellProduction}
k_{\rm p}=\kappa\rho-k_0\, ,
\end{equation}
where $\kappa$ and $k_0$ are two positive phenomenological constants. Within the spheroid, metabolites diffuse with a coefficient $D$ and are consumed or absorbed by the cells. Similarly, we assume a linear dependence of this absorption term in the metabolite concentration with a constant absorption rate $\alpha$, as it has been observed in avascular tumour growth~\cite{JiangMultiscale2005}:
\begin{equation}\label{eqNutrientDynamics}
\partial_t\rho=D\Delta\rho-\alpha\rho\, .
\end{equation}
Here, $\partial_t$ denotes the partial time derivative and $\Delta$ the Laplacian operator. Note that the convective term has been ignored in this equation, which is justified if nutrient diffusion is fast compared to its convective transport by the cell flow.

In the center of the spheroid $\mathbf{r}=\mathbf{0}$, the cell-velocity field vanishes and the metabolite concentration remains positive. At the outer surface, the metabolite concentration equals the external concentration $\rho^{\rm ext}$ and the cell velocity equals that of the interface. Labelling $R$ the spheroid stationary radius and $\delta R$ its perturbation, these conditions read $\rho_{r=R+\delta R}=\rho^{\rm ext}$ and
\begin{equation}\label{eqKinematicCondition}
\partial_t(\delta R)=\mathbf{v}_{r=R+\delta R}\cdot \mathbf{e}_r\, ,
\end{equation}
where $\mathbf{e}_r$ is the unit radial vector and $\mathbf{v}$ is evaluated at the perturbed interface location. Finally, the outer surface is subjected solely to the isotropic, external pressure. The tangential component $\sigma_{\rm nt}$ of the stress tensor therefore vanishes and its normal component is given by Laplace's law with surface tension $\gamma$: $(\sigma_{\rm nn})_{r=R+\delta R}=-P^{\rm ext}-\gamma H$, where $H$ is the local curvature. Note that the first boundary condition mentioned here, in defining a specific location for the center of the spheroid, breaks Galilean invariance. We comment in the following on the signification and consequences of this boundary condition.

\section{Stationary solution}

\subsection{Stationary equations}

The stationary equations are characterised by spherical symmetry with radial coordinate $r$. The stationary continuity equation~\ref{eqContinuity} and metabolite-diffusion equation~\ref{eqNutrientDynamics} reduce to:
\begin{equation}\label{eqContinuityStat}
\frac{1}{r^2}\,\partial_r(r^2\,v_r) = k_{\rm p}
\end{equation}
and
\begin{equation}\label{eqNutrientContinuityStat}
\frac{D}{r^2}\,\partial_r(r^2\partial_r\rho)=\alpha\rho\, ,
\end{equation}
together with the corresponding boundary conditions $(\rho^{\rm stat})_{r=R} = \rho^{\rm ext}$ and $(v_r^{\rm stat})_{r=R} = 0$. The stationary force-balance condition and expressions of the stress-tensor components are given in~\ref{appAddStatEq}.

\subsection{Stationary solutions}

To integrate this system of equations, we start by integrating the metabolite diffusion equation~\ref{eqNutrientContinuityStat}. With the characteristic metabolite-penetration length $l_D=\sqrt{D/\alpha}$ and the reduced variables $\bar{r}=r/l_D$ and $\bar{R}=R/l_D$, the metabolite concentration and the cell-production rate read
\begin{equation}\label{eqNutrientStatSolution}
\rho^{\rm stat}=\rho^{\rm ext}\frac{\bar R}{\sinh{\bar R}}\, \frac{\sinh{\bar r}}{\bar r}
\end{equation}
and
\begin{equation}\label{eqProdRateStatSolution}
k^{\rm stat}_{\rm p}=\kappa\rho^{\rm ext}\,\frac{\bar R}{\sinh{\bar R}}\, \frac{\sinh{\bar r}}{\bar r}-k_0\, .
\end{equation}
We can see from Eq.~\ref{eqProdRateStatSolution} that a stationary solution with spherical symmetry exists as long as $\kappa\rho^{\rm ext}>k_0$ to have a positive cell-division rate at the outer rim of the spheroid. Finally, the cell-velocity field reads
\begin{equation}\label{eqVelocityStatSolution}
v^{\rm stat}_r=\kappa\rho^{\rm ext}\,\frac{R}{\sinh{\bar R}}\left[\frac{\cosh{\bar r}}{\bar r}-\frac{\sinh{\bar r}}{\bar r^2}\right]-\frac{1}{3}k_0r\, .
\end{equation}
The boundary condition $(v_r^{\rm stat})_{r=R} = 0$ then leads to the following equation for the stationary radius: 
\begin{equation}\label{eqStatRadius}
\frac{1}{\bar{R}}\coth{\bar{R}}-\frac{1}{\bar{R}^2}=\frac{k_0}{3\kappa\rho^{\rm ext}}\, .
\end{equation}
Note that the product $\kappa\rho^{\rm ext}$ is linked to $k_0$ and the cell-production rate at the outer rim $k_{\rm p}^{\rm ext}$ by $k_{\rm p}^{\rm ext}=\kappa\rho^{\rm ext}-k_0$. The complete expressions of all the other quantities in this stationary state are given in~\ref{appAddStatSolutions}.

\section{Mode computation}\label{secModes}

\subsection{Perturbed axisymmetric equations}

We now investigate the linear stability of this stationary solution to axisymmetric perturbations. We choose a system of coordinates composed of radial coordinate $r$, polar angle $\theta$, and azimuthal angle $\phi$, and we study the perturbations with axial symmetry around $\theta=0$. The model equations with axial symmetry read
\begin{equation}\label{eqContinuityAxisym}
\frac{1}{r^2}\,\partial_r(r^2\,v_r) + \frac{1}{r\sin{\theta}}\,\partial_\theta(\sin{\theta}\,v_\theta)= k_{\rm p}
\end{equation}
for the continuity equation~\ref{eqContinuity} and
\begin{equation}\label{eqNutrientDynamicsAxisym}
\partial_t\rho=\frac{D}{r^2}\,\partial_r(r^2\partial_r\rho)+\frac{D}{r^2\sin{\theta}}\,\partial_\theta(\sin{\theta}\partial_\theta\rho)-\alpha\rho
\end{equation}
for the metabolite diffusion equation~\ref{eqNutrientDynamics}. Contrary to the stationary system of equations, the cell-velocity field cannot be solved independently of the force-balance condition~\ref{eqForceBalance}. The other, coupled equations, are given in~\ref{appAddPerturbedEq}.

\subsection{Linear decomposition}

To integrate this system of equations to linear order in perturbations, we expand the angular dependence of the different perturbative fields onto the basis of axisymmetric, spherical harmonics. Following ref.~\cite{HappelLow2012}, the perturbations $\delta R$, $\delta v_r$, $\delta P$, and $\delta \rho$ are expanded onto the basis of Legendre polynomials $\left(P_n(\cos{\theta})\right)_{n\in\mathbb{N}}$, and the perturbation $\delta v_\theta$ is expanded onto the basis of the Gegenbauer polynomials $\left(I_n(\cos{\theta})\right)_{n\in\mathbb{N}}$, where $I_n=(P_{n-1}-P_{n+1})/(2n+1)$. The components $v_r^{(n)}$, $P^{(n)}$, $\rho^{(n)}$, and $v_\theta^{(n)}$ of respectively $\delta v_r$, $\delta P$, $\delta \rho$, and $\delta v_\theta$ under this expansion are functions of the radial coordinate $r$, as the components $R^{(n)}$ of the expansion of the interface location $\delta R$ are simple numbers. Explicitly, the $\delta v_r$ expansion for example reads
\begin{equation}
\delta v_r(r,\theta)=\sum_{n=0}^\infty v_r^{(n)}\,P_n(\cos{\theta})\, ,
\end{equation}
as the $\delta v_\theta$ expansion reads
\begin{equation}
\delta v_\theta(r,\theta)=\sum_{n=1}^\infty v_\theta^{(n)}\,\frac{I_{n+1}(\cos{\theta})}{\sin{\theta}}\, ,
\end{equation}
where the sum starts at $n=1$ since the mode $n=0$ is purely radial. Using these expansions, the metabolite diffusion equation~\ref{eqNutrientDynamics} leads to
\begin{equation}\label{eqNutrientDynamicsPerturbed}
\partial_r^2\rho^{(n)}+\frac{2}{r}\,\partial_r\rho^{(n)}-\left[\frac{\alpha+\omega_n}{D}+\frac{n(n+1)}{r^2}\right]\rho^{(n)}=0\, ,
\end{equation}
where $\rho^{(n)}$ is the component for the mode number $n$ of the nutrient field $\rho$ and $\omega_n$ the corresponding growth rate. The components of the perturbed velocity field are then determined by the perturbed continuity equation
\begin{equation}\label{eqContinuityPerturbed}
\partial_r\,v_r^{(n)} + \frac{2}{r}\,v_r^{(n)} + \frac{1}{r}\,v_\theta^{(n)} = \kappa\,\rho^{(n)}
\end{equation}
as well as by the force-balance equations, further given in ~\ref{appAddModeDecomp}.

\subsection{Explicit solution in the limit of fast diffusion}\label{sseExplicitSol}

To solve equation~\ref{eqNutrientDynamicsPerturbed}, we first consider the regime where the relaxation or growth of the perturbation modes as well as metabolite consumption are slow compared with metabolite diffusion over the characteristic lengths involved, at most equal to the spheroid radius. In that limit, the term $[(\alpha+\omega_n)/D]\rho^{(n)}$ can be neglected in front of $[n(n+1)/r^2)]\rho^{(n)}$. This approximation is certainly not valid for the mode $n=0$, which needs to be computed separately. The calculation happens to be singular as well for $n=1$. For $n\geq 2$, the solution of equation~\ref{eqNutrientDynamicsPerturbed} in this approximation is a simple power law, and we can further obtain the other perturbed quantities analytically. We have, for all $n\geq 2$:
\begin{align}\label{eqPerturbedNutrientVelocity}
\rho^{(n)} &= c_n\bar{r}^n\nonumber\\
v_r^{(n)} &= a_n\bar{r}^{n+1}+b_n\bar{r}^{n-1}\nonumber\\
v_\theta^{(n)} &= \left[-(n+3)a_n+\kappa\,R\,c_n\right]\bar{r}^{n+1}-(n+1)b_n\bar{r}^{n-1}\, ,
\end{align}
where $a_n$, $b_n$, and $c_n$ are three integration constants, all proportional to the amplitude $R^{(n)}$ of the perturbed radius. The other quantities can be further expressed as linear combinations of these three integration constants. The other obtained expressions are given in~\ref{appAddModeExpression}.

The integration contants are then determined using the boundary conditions. Plugging these solutions into the kinematic equation~\ref{eqKinematicCondition}, we finally get the following mode growth rates:
\begin{equation}\label{eqModeZerothOrder}
\omega_n=\frac{2n^2+5n+3}{2n^2+4n+3}\,k_{\rm p}^{\rm ext}-\frac{1}{2n^2+4n+3}\left[\frac{n+1}{3}\,k_0\left(\frac{R}{l_D}\right)^2+(2n^2+5n+2)\frac{\gamma}{2\eta}\,\frac{n}{R}\right]
\end{equation}
for all $n\geq 2$. For the modes $n=0$ and $n=1$, we get separately $\omega_0=k_{\rm p}^{\rm ext}-(1/9)k_0\bar{R}^2$ and $\omega_1=(1/3)k_{\rm p}^{\rm ext}-(1/18)k_0\bar{R}^2$. These latter expressions are independent of the surface tension $\gamma$. This is because, for these two modes, perturbations in the curvature occur only to second order or higher.

\section{Results}\label{secResults}

We can now discuss the instability in this fast-diffusion regime. The first term in equation~\ref{eqModeZerothOrder} is destabilising and proportional to the cell-production rate at the outer surface. The second term is stabilising and results for one part from cell-death processes controlled by the parameters $\alpha/D$ (via $l_D$) and $k_0$, and from the other part by surface tension. Increasing the viscosity lowers the contribution of surface tension, destabilising the spheroid. This underlines the mechanical origin of the instability, which relies on internal viscous stresses within the tissue, generated by differential cell flows (see figure~\ref{figSchematic}). Therefore, this instability can only exist around a kinematic steady state with nonzero permanent cell flows, here from the outer surface toward the spheroid core.

Considering the stationary condition equation~\ref{eqStatRadius}, we can verify that the modes $n=0$ and $n=1$ are always stable (see~\ref{appAddModeExpression}, equation~\ref{eqModes0and1ZerothOrderBis}). For the other modes, in the absence of surface tension, the first term in equation~\ref{eqModeZerothOrder} is dominant at large $n$, meaning that, without this contribution, the spheroid is always unstable with a rate asymptotically equal to that of cell division at its outer rim. Surface tension however stabilises the spheroid, since it contributes by a term scaling as $\gamma q_n/(2\eta)$ at large $n$, where $q_n=n/R$. Depending on the values of the different parameters, we therefore expect a potential instability to develop at a finite value of $n$, corresponding to a finite wavelength $\lambda_n\sim 2\pi R/n$. 

It is interesting to investigate the behaviour of the most unstable mode as a function of the stationary radius $R$. Since the equation characterising this mode is in general fourth order in $n$, we investigate separately the limits of small and large spheroids. In the limit of small radii, the least stable mode is $\omega_0$, since curvature is large and surface tension strongly stabilises all modes for larger values of $n$. In the limit of large radii, the most unstable mode occurs at $n=n_{\rm max}$ with the following asymptotic expression, linear in $R$: $n_{\rm max} \simeq \sqrt{\eta\kappa\rho^{\rm ext}/(\gamma l_D)}\,R$. This scaling indicates that the associated wavelength converges toward a finite value at large radius $R$. In this limit, we asymptotically reach the case of a flat surface, with an instability that evokes what has been proposed for epithelial tissues~\cite{BasanUndulation2011,RislerMorphological2013}. The corresponding growth rate reads $\omega_{\rm max} \simeq \kappa\rho^{\rm ext}-\sqrt{\gamma\kappa\rho^{\rm ext}/(\eta l_D)}$.

In the generic case where metabolite diffusion is not necessarily fast compared to metabolite consumption or perturbation growth, the solution for $\rho^{(n)}$ in equation~\ref{eqNutrientDynamicsPerturbed} is a function of the associated growth rate $\omega_n$. Equation~\ref{eqKinematicCondition} then becomes an implicit equation for the growth rate $\omega_n$, which cannot be solved analytically. We report the implicit equation corresponding to the mode $n=0$ as an example in~\ref{appImplicitEquation}, equation~\ref{eqMode0ExactComputation}.

To compute the growth rates $\omega_n$ numerically, we now estimate the different parameter values. The shear viscosity of cellular aggregates has been estimated in different experiments, leading to $\eta\simeq 10^4-10^5$~Pa$\cdot$s~\cite{ForgacsViscoelastic1998,MarmottantRole2009,GuevorkianAspiration2010}. Tissue surface tensions have been measured for different tissue types. Measurements for $\gamma$ give values ranging from a fraction up to several millinewton per meter~\cite{ForgacsViscoelastic1998,SchotzQuantitative2008,MgharbelMeasuring2009,GuevorkianAspiration2010}. We further assume a typical cell-division rate at the outer surface of the spheroid $k_{\rm p}^{\rm ext}$ and a cell-death rate in the absence of metabolites $k_0$ of one per day. Cellular growth within the spheroid can be limited by different types of metabolites. Depending typically on the molecular size of a given metabolite, its diffusion coefficient can take different values. Estimates of the diffusion coefficient of growth factors and glucose in avascular tumours range from 10$^{-6}$~cm$^2\cdot$h$^{-1}$ for growth factors~\cite{JiangMultiscale2005} to 1.5~10$^{-3}$~cm$^2\cdot$h$^{-1}$ for glucose~\cite{FreyerDetermination1983,CasciariGlucose1988,JiangMultiscale2005}, or even larger values for oxygen~\cite{Mueller-KlieserMethod1984,JiangMultiscale2005}. In the following, we shall investigate the influence of a variation of this particular parameter. Finally, to obtain radii of a few hundred micrometres, we choose to have comparable values for the characteristic penetration length of metabolites $l_D$. This leads to values for the metabolite-consumption rate $\alpha$ of several tenths per day for $D\sim$10$^{-6}$~cm$^2\cdot$h$^{-1}$, scaled accordingly when $D$ is varied to keep $l_D$ constant.

\begin{figure}[t]
\begin{center}
\includegraphics[width=0.75\textwidth]{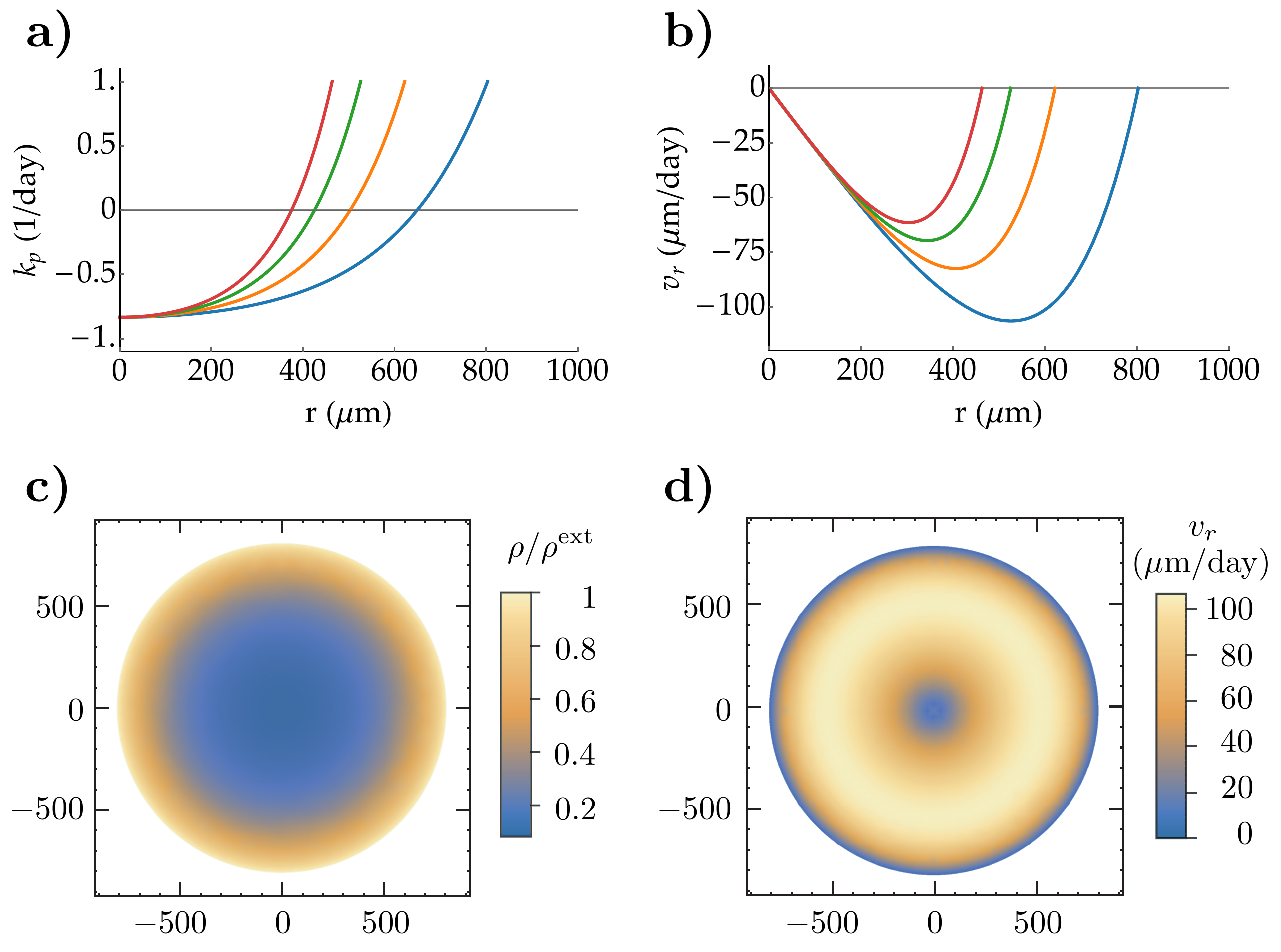}
\end{center}
\caption
{\label{figStatResults} Stationary state of a multicellular spheroid. (a,b) The cell-production rate $k_{\rm p}$ (a) and the radial cell-velocity field $v_r$ (b) are displayed as functions of the radial distance $r$ for four different values of the stationary radius $R$. Parameters common to the four sets of curves are: $\eta=80$~kPa$\cdot$s, $\gamma=100$~$\mu$N$\cdot$m$^{-1}$, and $k_{\rm p}^{\rm ext} = k_0 = 1$~d$^{-1}$ (d$^{-1}$ stems for ``per day''). The stationary state depends on the diffusion coefficient of metabolites $D$ and metabolite-absorption rate $\alpha$ through their ratio $D/\alpha=l_D^2$ only. We display the resulting stationary state for four different values of this ratio. The corresponding stationary radii and metabolite penetration lengths are $R\simeq$~803~$\mu$m and $l_D\simeq$~170~$\mu$m (blue curves), $R\simeq$~622~$\mu$m and $l_D\simeq$~131~$\mu$m (orange curves), $R\simeq$~526~$\mu$m and $l_D\simeq$~111~$\mu$m (green curves), and $R\simeq$~464~$\mu$m and $l_D\simeq$~98~$\mu$m (red curves). (c,d) To give a better feeling of the spherical symmetry, we display in panel (c) the stationary metabolite-concentration profile with respect to its outer-surface value $\rho/\rho^{\rm ext}$ and in panel (d) the corresponding radial cell-velocity field $v_r$ (in $\mu$m$\cdot$d$^{-1}$), for the parameter set corresponding to the blue curves of panels (a) and (b). Plots are made in a plane of symmetry of the spheroid.}
\end{figure}

We illustrate in figure~\ref{figStatResults} the obtained results for the stationary solution of the model.
The stationary-state profiles show that cells divide preferentially close to the outer surface and disappear in the centre (panel~(a)), due to the lack of metabolites penetrating the tissue (panel~(c)).  As a result, cells flow inwards from the outer surface to the centre (panels~(b) and (d)). This result is in agreement with experimental measurements of cellular flows in multicellular spheroids using fluorescently labeled particles~\cite{DelarueMechanical2013}.

\begin{figure}[t]
\begin{center}
\includegraphics[width=0.75\textwidth]{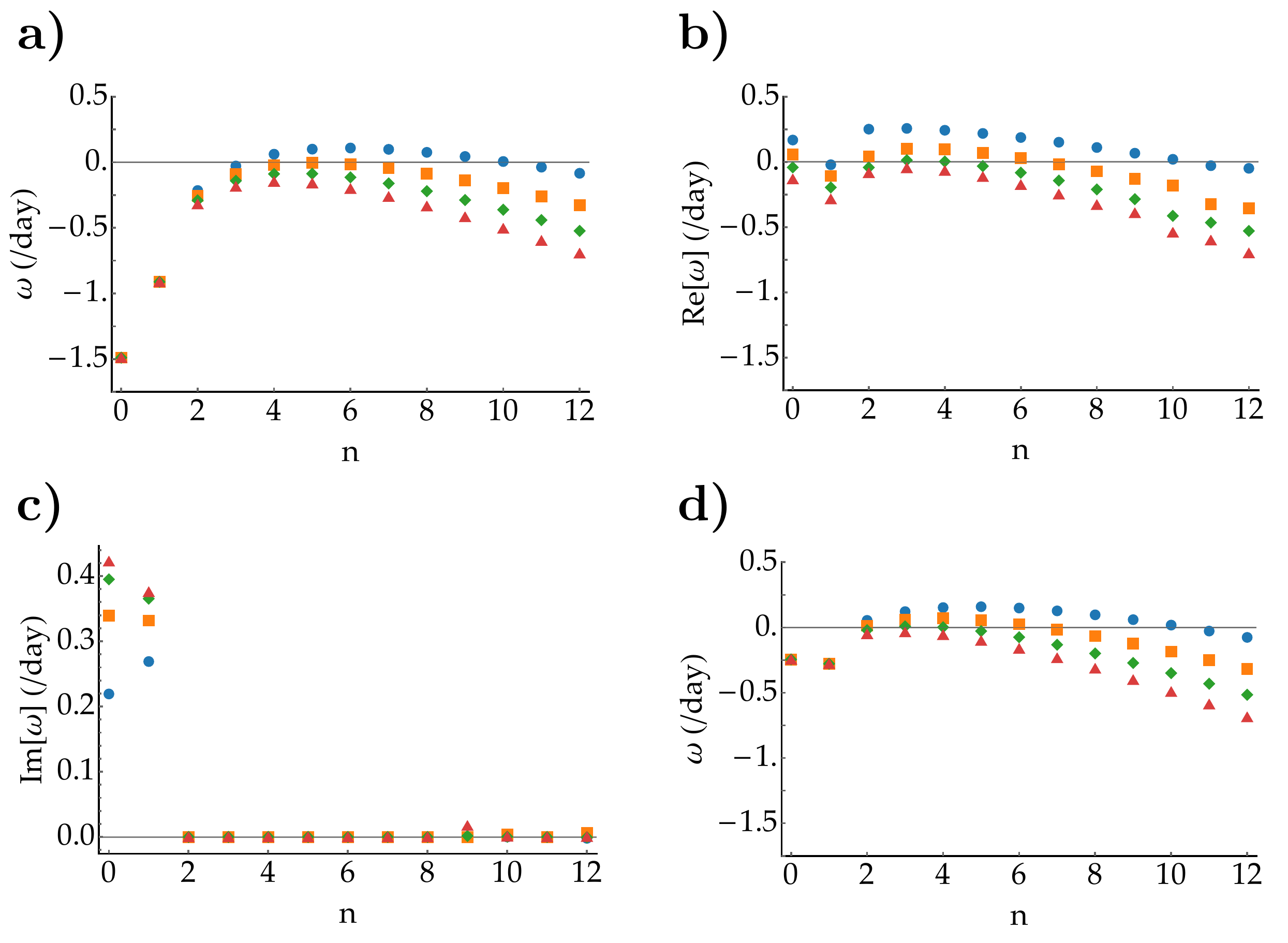}
\end{center}
\caption
{\label{figModeResults} Viscocapillary mode growth rates of a multicellular spheroid as functions of the mode number $n$, obtained (a) in the analytic limit of section~\ref{sseExplicitSol}, (b,c,d) numerically using the full model in both limits of slow (b,c) and fast (d) diffusion. (a) Parameters are the same as those of figure~\ref{figStatResults} using the same colour code. The mode-growth rates depend on the diffusion coefficient of metabolites $D$ and metabolite-absorption rate $\alpha$ through their ratio $D/\alpha=l_D^2$ only. We have $l_D\simeq$~170~$\mu$m (blue circles), $l_D\simeq$~131~$\mu$m (orange squares), $l_D\simeq$~111~$\mu$m (green diamonds), and $l_D\simeq$~98~$\mu$m (red triangles). (b,c) Real (b) and imaginary (c) parts of the mode-growth rates computed numerically with a diffusion coefficient of metabolites equal to that of growth factors: $D=5\cdot 10^{-10}$~cm$^2\cdot$s$^{-1}$~\cite{JiangMultiscale2005}. The metabolite-absorption rate is varied accordingly to keep the same steady state as in figure~\ref{figStatResults}: $\alpha=0.15$~d$^{-1}$ (blue circles), $\alpha=0.25$~d$^{-1}$ (orange squares), $\alpha=0.35$~d$^{-1}$ (green diamonds), and $\alpha=0.45$~d$^{-1}$ (red triangles). (d) Mode-growth rates computed numerically with a diffusion coefficient of metabolite equal to that of glucose: $D=5\cdot 10^{-7}$~cm$^2\cdot$s$^{-1}$~\cite{FreyerDetermination1983,CasciariGlucose1988,JiangMultiscale2005}. The metabolite-absorption rate is scaled accordingly by a factor of a thousand to perturb around the same steady state as previously: $\alpha=150$~d$^{-1}$ (blue circles), $\alpha=250$~d$^{-1}$ (orange squares), $\alpha=350$~d$^{-1}$ (green diamonds), $\alpha=450$~d$^{-1}$ (red triangles). The mode-growth rates in (a) and (d) are real.}
\end{figure}

We illustrate in figure~\ref{figModeResults} the central result of our study, that is the obtained mode structure of the instability, for the four different steady states presented in figure~\ref{figStatResults}a and b, using the same colour code.
The system is unstable as soon as at least one perturbation mode displays a growth rate $\omega$ with a positive real part. When this is the case, we expect the fastest growing modes to develop first, corresponding to the maximum of each series of points presented in these plots. A visual display of the shapes associated with the deformation modes $n=0$ to 5 is shown in~\ref{appModeShapes}, figure~\ref{figFourierModes}. Figure~\ref{figModeResults}a shows the mode structure in the approximation of the analytic solution of section~\ref{sseExplicitSol}. There is a transition at finite wavelength as a function of the metabolite-consumption rate $\alpha$ from a stable to an unstable regime, with a range of unstable modes. These appear for example when nutrient consumption decreases at fixed external cell-division rate, surface tension and viscosity. With the parameters chosen here, the first mode to become unstable is $n=5$ (figure~\ref{figModeResults}a, orange squares). This is associated with a stationary radius $R_5\simeq$~622~$\mu$m and corresponds to an unstable wavelength $\lambda_5\simeq$~782~$\mu$m.

In figure~\ref{figModeResults}b and c, we display respectively the real and imaginary parts of the mode growth rates obtained in the full model solved numerically, with the parameter sets of figure~\ref{figStatResults} and a diffusion coefficient $D=5\cdot 10^{-10}$~cm$^2\cdot$s$^{-1}$, corresponding approximately to that estimated for growth factors~\cite{JiangMultiscale2005}. The system also displays first an instability at finite wavelength (for the mode $n=3$, green diamonds), but the modes $n=0$ and $n=1$ can now be unstable. In addition, these instabilities are oscillatory, as characterised by non-zero imaginary parts of $\omega$, corresponding to the characteristic frequencies of the unstable oscillations (see figure~\ref{figModeResults}c). This result stems from the existence of a delay in the response of the cells located in the bulk of the aggregate to a given perturbation of the outer surface, due to the finite kinetics of metabolite penetration into the spheroid.

Our model ignores the convective cellular flow in the metabolite diffusion equation~\ref{eqNutrientDynamics}. For this approximation to be valid, one requires that $D/l_D$ be larger than the amplitude~$v$ of the cellular flow within the aggregate. Figure~\ref{figModeResults}d displays the mode growth rates obtained with a diffusion coefficient $D$ a thousand times that of figure~\ref{figModeResults}b,c, corresponding approximately to the diffusion coefficient of small nutrient molecules such as glucose~\cite{FreyerDetermination1983,CasciariGlucose1988,JiangMultiscale2005}. With this value of the diffusion coefficient, we have $(\ell_D\cdot v)/D\sim 4.10^{-3}\ll 1$ when $v$ is estimated from the curves shown in figure~\ref{figStatResults}b in the least favorable case, satisfying largely the required condition for neglecting convective flows. We show in figure~\ref{figModeResults}d that, perturbing around the same steady state by rescaling the metabolite-absorbing rate $\alpha$, the instability occurs at a similar radius and finite wavelength as those reported above. In addition, the growth rates of the high-order modes ($n\geq 5$--6) are largely unchanged. 
Interestingly, the oscillatory instability is lost for $n=0$ and $n=1$. This is the signature of the fact that here metabolite diffusion is sufficiently fast to allow for an almost instantaneous response of the inner cells to perturbations of the outer surface. We however do not recover the analytic results of figure~\ref{figModeResults}a for small values of $n$. This stems from the fact that, in the results of figure~\ref{figModeResults}d, we scale the metabolite consumption rate with the diffusion coefficient to keep the same steady-state sizes as in figure~\ref{figModeResults}b, as the analytic limit was obtained for small values of $\alpha/D$ with respect to $1/R^2$. We show in \ref{appDiffusionStudy}, figure~\ref{figStudyDiffusion}, the mode structure obtained with intermediate values of the diffusion coefficient $D$ between those of figures~\ref{figModeResults}b,c and \ref{figModeResults}d, following the same rescaling procedure of the metabolite-consumption rate $\alpha$.

As mentioned at the end of section~\ref{secModelDesc}, our boundary conditions specify that the cell-velocity field vanishes at the center of the spheroid located at $\mathbf{r}=\mathbf{0}$, which breaks Galilean invariance. Doing so, the mode $n=1$ here corresponds to an actual deformation of the flow pattern within the spheroid, with an outer boundary displaced with respect to the point where the flow pattern converges. As a consequence, this mode is not necessarily marginal, contrary to many standard spherical-harmonic perturbation analyses. A similar interpretation of the mode $n=1$ is found, e.g., in the context of the deformations of the actin cortex of a spherical cell~\cite{SalbreuxShape2007}, where it corresponds to a relative translation of the inner part of the cell cortex with respect to the outer boundary of the cell, rather than to a global translation of the whole system. Therefore, even if this mode corresponds to a global translation of the inner boundary of the cortex with no deformation, the relative positions of the inner and outer boundaries of the cortex do vary, which corresponds to a spatial variation of its overall thickness. As a result, the mode $n=1$ is not marginal. Similarly, here, we investigate relative displacements of the different cell layers with each other. Each cell layer is purely translated in the mode $n=1$, but the relative positions of the different layers varies within the spheroid. Therefore, even though the overall external shape of the spheroid is unchanged, the cell-flow pattern is modified by this perturbation.

At threshold, the unstable modes are expected to grow exponentially in time, albeit potentially on long timescales. If this scenario is expected at sufficiently small amplitudes, where the linear regime of small deformations around the spherical shape is valid, large-amplitude deformations are expected to follow another, more complicated dynamics. In particular, we expect that when perturbation amplitudes reach a finite fraction of the spheroid radius, geometric nonlinearities will induce more complex flow patterns because of an asymmetry between inward and outward deformations. We expect eventually that outward protrusions would at late stages outgrow inward protrusions, the latter being limited by the original spheroid size. We can speculate that such an unbalance would lead to a global growth of the overall spheroid mass.

\section{Discussion}\label{secDiscussion}

In this work, we have shown the potential existence of an instability in spherical tissues, which can develop from steady states that are limited by the supply of metabolites diffusing from their microenvironment. The present instability relies neither on cell motility nor on the presence of external forces, but rather stems from the presence of viscous shear stresses generated by the spatial organisation of cell renewal within the tissue. We have shown that the instability develops at a finite wavenumber, which reflects a balance of viscous shear stresses with those stemming from surface tension. We propose that this mechanism could be observed in multicellular spheroids in culture, which would be an ideal system for testing the influence of different parameters controlling the instability, such as tissue viscosity, surface tension, and metabolite supply. The former two could be changed, e.g., by varying the expression of proteins implicated in cell-surface adhesion or actin-cortex contractility~\cite{LecuitCell2007,SchotzQuantitative2008,ManningCoaction2010}.

The proposed instability here evokes other already proposed instabilities in the context of the cell cytoskeleton, driven by actin-polymerisation dynamics~\cite{Callan-JonesViscousFingeringLike2008,Blanch-MercaderSpontaneous2013}. In these studies of the stability of cell fragments on a substrate, an inward flow is driven by actin polymerisation at the outer edge and actin depolymerisation in the bulk. As a result, an originally circular cell fragment can become unstable and spontaneously acquire a polarisation. Important differences between our current study and these previous works however exist. In the stability analysis of circular cell fragments, the generation of new material occurs only at the outer surface, where actin polymerises. In our current model, cell production is a global, bulk effect, which varies continuously within the spheroid. The second difference is the rheology, which here corresponds a Stoke flow with no contact with an external substrate, as in the case of cell fragments there is a Darcy flow, rendering the two types of instability different. Associated to that, the third difference is that our current study is three dimensional as these previous works are two dimensional. As a result, our flow pattern has no anchoring to the external world and requires a minimal thickness over which cells divide to become unstable.

In addition to be applicable to multicellular aggregates, one can wonder if similar instabilities could arise {\it in vivo}. During development, transition from solid-like to fluid-like tissue properties, tissue surface tension, and flow patterns have been shown to play a crucial role (see, e.g., \cite{MongeraFluidtosolid2018}). Recently, three-dimensional aggregates of mouse embryonic stem cells have been shown to undergo a first morphological transformation from a spherical into an oblong shape during gastrulation, associated with a reduced level of E-cadherin expression at the developing tip~\cite{HashmiCellstate2020}. Such shapes resemble a superposition of instability modes such as $n=2$ and 3, and potentially higher, as illustrated in figure~\ref{figFourierModes}. Interestingly with respect to our current study, the polarisation of E-cadherin expression precedes the onset of tip formation, and when the level of E-cadherin expression is maintained high, the aggregate remains generally devoid of any pole~\cite{HashmiCellstate2020}. These observations suggest a role for a reduced surface tension in the development of the protrusion, similar to what we are proposing here. However, in these examples, and to our knowledge more generally during development, it seems that an original inhomogeneity in the tissue rheological parameters---such as surface tension and viscosity---is at the origin of the shape formation. Such inhomogeneities are however absent in our proposed mechanism, which relies solely on the presence of a permanent flow of duplicating cells.

A domain to which we can speculate that the present mechanism applies is the evolution of microtumours after a long period of dormancy. Small primary tumours or early metastases often enter a state where their sizes remain steady, before they resume growth or disappear~\cite{Aguirre-GhisoModels2007}. Such a dormant state can last for a long time and is at the origin of late cancer reappearance, sometimes years after the original treatment~\cite{UhrCancer1997,KarrisonDormancy1999}. It is recognised that microscopic, clinically occult tumours are very common in the population and that only a tiny fraction of them ever becomes clinically relevant~\cite{BlackAdvances1993,FolkmanCancer2004,NaumovRole2006}. Understanding the factors that can destabilise a dormant tumour is therefore of crucial importance. The main mechanisms at the origin of such steady states are angiogenic dormancy, cellular dormancy (G0-G1 arrest) and immunosurveillance~\cite{Aguirre-GhisoModels2007}. In angiogenic dormancy, the tumour is limited in its growth by the lack of 
metabolites, which are brought by blood vessels that do not penetrate the tumour~\cite{FolkmanCancer2004,NaumovRole2006,Aguirre-GhisoModels2007}. This limitation keeps the microtumour to sizes typically smaller than 1--2~mm in diameter, until the angiogenic switch is triggered~\cite{HanahanPatterns1996,SemenzaAngiogenesis2003,NaumovRole2006}.

Tumour-growth models have explored the effects of a wide variety of biological processes~\cite{PreziosiCancer2003,TracquiBiophysical2009,ByrneDissecting2010,RislerFocus2015,ScottCancer2015,PortaPhysics2017}. In support of the current surface mechanism, it has recently been shown that colon cancer xenografts grow primarily from their surfaces~\cite{LamprechtMulticolor2017,LenosStem2018}, which corresponds to the steady-state patterns of cell duplications on which our current study relies. In addition, clonal expansion largely depends on the location of a clone within the tumour~\cite{HeijdenSpatiotemporal2019}, suggesting that differences in geometrical or physical properties within the tumour are major contributors to heterogeneous clonal expansion. This latter observation leads us to speculate that, after a first instability such as the one proposed here, the resulting irregular shape creates different microenvironments for different parts of the tumour, further driving different epigenetic and maybe even later genetic transformations by diverse selection processes.

The proposed instability can be triggered by a change of internal properties such as cell-cell adhesion strength, cell-renewal rate, metabolite supply, or other parameters affecting the overall spheroid size. In a microtumour, such changes might be multifactorial, e.g. ageing, a change in the person's metabolism, in the immune system's activity, or in drug delivery or efficiency. While our model does not address the long-time evolution of these parameters, a multicellular aggregate or a microtumour can still be unstable by the mechanism proposed here. Our study could therefore be of importance for determining which parameters control the spherical stability. Using multicellular spheroids as model systems, it could on the long run participate in guiding which aspects of tumour development should be targeted by medication.

\ack
We thank F.~Brochard-Wyart, D.~Gonzalez-Rodriguez, K.~Guevorkian, J.-F. Joanny, and J.~Prost for insightful discussions and useful comments on the manuscript. This work received support from the LabEx Cell(n)Scale (former CelTisPhyBio), grants ANR-11-LABX-0038 and ANR-10-IDEX-0001-02.

\appendix

\section{Additional stationary mechanical equations}\label{appAddStatEq}

The stationary, non-trivial component of the force-balance equation~\ref{eqForceBalance} reads
\begin{equation}\label{eqForceBalanceStat}
\partial_r\sigma_{rr}+\frac{2}{r}(\sigma_{rr}-\sigma_{\theta\theta})= 0\, ,
\end{equation}
and the non-zero components of the dynamic part of the stress tensor $\sigma'$ as given by equation~\ref{eqDeviatoricStress} read
\begin{align}\label{eqDeviatoricStressStat}
\sigma'_{rr}&=2\eta\,\partial_r v_r +\left(\zeta-\frac{2}{3}\eta\right)\nabla\cdot\mathbf{v}\nonumber\\
\sigma'_{\theta\theta}&=\sigma'_{\phi\phi}=2\eta\,\frac{v_r}{r}+\left(\zeta-\frac{2}{3}\eta\right)\nabla\cdot\mathbf{v}\, ,
\end{align}
where 
\begin{equation}\label{eqDivergenceStat}
\nabla\cdot\mathbf{v}=\partial_r v_r+\frac{2}{r}v_r\, .
\end{equation}
The corresponding boundary condition reads
\begin{equation}\label{eqBoundaryCondStressStat}
(\sigma_{rr}^{\rm stat})_{r=R} = -P^{\rm ext}-\frac{2\gamma}{R}\, .
\end{equation}

\section{Additional stationary expressions}\label{appAddStatSolutions}

The non-trivial components of the dynamic part of the stress tensor read
\begin{equation}\label{eqDeviatoricStressSolutionStatLin1}
\sigma'^{\, \rm stat}_{rr}=\kappa\rho^{\rm ext}\,\frac{\bar R}{\sinh{\bar R}}\left[-4\eta\frac{\cosh{\bar r}}{\bar r^2}+\left(\zeta+\frac{4}{3}\eta+\frac{4\eta}{\bar r^2}\right)\frac{\sinh{\bar r}}{\bar r}\right]-\zeta k_0
\end{equation}
and
\begin{equation}\label{eqDeviatoricStressSolutionStatLin2}
\sigma'^{\, \rm stat}_{\theta\theta}=\kappa\rho^{\rm ext}\,\frac{\bar R}{\sinh{\bar R}}\left[2\eta\frac{\cosh{\bar r}}{\bar r^2}+\left(\zeta-\frac{2}{3}\eta-\frac{2\eta}{\bar r^2}\right)\frac{\sinh{\bar r}}{\bar r}\right]-\zeta k_0\, .
\end{equation}
The pressure is given by
\begin{equation}\label{eqPressureSolutionStat1}
P^{\, \rm stat}=P^{\, \rm stat}_{\bar r=0}+\kappa\rho^{\rm ext}\,\left(\zeta+\frac{4}{3}\eta\right)\,\frac{\bar R}{\sinh(\bar R)}\left(\frac{\sinh(\bar r)}{\bar r}-1\right)\, ,
\end{equation}
where
\begin{equation}\label{eqPressureConstantStat1}
P^{\, \rm stat}_{\bar r=0}=P^{\rm ext}+\frac{2\gamma}{R}-\zeta k_0+\frac{\kappa\rho^{\rm ext}}{\bar R^2}\left[4\eta\left(1-\bar R\coth{\bar R}\right)+\left(\zeta+\frac{4}{3}\eta\right)\frac{{\bar R}^3}{\sinh{\bar R}}\right]\, .
\end{equation}

\section{Additional perturbed axisymmetric equations}\label{appAddPerturbedEq}

With axisymmetry, the non-trivial components of the force-balance equation~\ref{eqForceBalance} read
\begin{align}\label{eqForceBalanceAxisym}
\partial_r\sigma_{rr}+\frac{1}{r}\partial_\theta\sigma_{r\theta}+\frac{1}{r}\left[2\sigma_{rr}-\sigma_{\theta\theta}-\sigma_{\phi\phi}+\sigma_{r\theta}\cot{\theta}\right]&=0\nonumber\\
\partial_r\sigma_{r\theta}+\frac{1}{r}\partial_\theta\sigma_{\theta\theta}+\frac{1}{r}\left[3\sigma_{r\theta}+(\sigma_{\theta\theta}-\sigma_{\phi\phi})\cot{\theta}\right]&=0\, ,
\end{align}
and the non-zero components of the dynamic part of the stress tensor, as given by equation~\ref{eqDeviatoricStress}, are
\begin{align}\label{eqDeviatoricStressAxisym}
\sigma'_{rr}&=2\eta\,\partial_r v_r+\left(\zeta-\frac{2}{3}\eta\right)\nabla\cdot\mathbf{v}\nonumber\\
\sigma'_{\theta\theta}&=\frac{2\eta}{r}(v_r+\partial_\theta v_\theta)+\left(\zeta-\frac{2}{3}\eta\right)\nabla\cdot\mathbf{v}\nonumber\\
\sigma'_{\phi\phi}&=\frac{2\eta}{r}(v_r+v_\theta\cot{\theta})+\left(\zeta-\frac{2}{3}\eta\right)\nabla\cdot\mathbf{v}\nonumber\\
\sigma'_{r\theta}=\sigma'_{\theta r}&=\frac{\eta}{r}\left(\partial_\theta v_r+r\partial_r v_\theta-v_\theta\right)\, ,
\end{align}
where 
\begin{equation}\label{eqDivergenceAxisym}
\nabla\cdot\mathbf{v}=\partial_r v_r+\frac{2}{r}v_r+\frac{1}{r\sin{\theta}}\partial_\theta(\sin{\theta}\,v_\theta)\, .
\end{equation}

\section{Additional perturbed axisymmetric mode decomposition}\label{appAddModeDecomp}

The components of the perturbed velocity field are determined by the perturbed continuity equation~\ref{eqContinuityPerturbed} as well as by the force-balance equations, which reduce to
\begin{align}\label{eqForceBalancePerturbed}
\left(2\partial_r^2+\frac{4}{r}\partial_r-\frac{4+n(n+1)}{r^2}\right)\,v_r^{(n)}+\left(-\frac{3}{r^2}+\frac{1}{r}\partial_r\right)\,v_\theta^{(n)}&=\frac{1}{\eta}\partial_r\bar{P}^{(n)}\nonumber\\
\left(r\partial_r^2+2\partial_r-\frac{2n(n+1)}{r}\right)\,v_\theta^{(n)}-n(n+1)\left(\frac{4}{r}+\partial_r\right)\,v_r^{(n)}&=-\frac{n(n+1)}{\eta}\bar{P}^{(n)}\, ,
\end{align}
where $\bar{P}^{(n)}=P^{(n)}-\left(\zeta-(2/3)\eta\right)\kappa\rho^{(n)}$.

To express the boundary conditions with axisymmetry, we need to express the local curvature $H$ at the outer surface of the spheroid. To first order in perturbations in $\delta R$, it is given by
\begin{equation}
H = \frac{2}{R}-[\partial_\theta^2+(\cot{\theta})\partial_\theta+2]\frac{\delta R}{R^2}\, .
\end{equation}
The different quantities evaluated at $r=R+\delta R$ read, to first order in perturbations:
\begin{align}\label{eqPerturbedBoundaryFields}
(\sigma_{\rm nn})_{r=R+\delta R} &= \sigma_{rr}^{\rm stat}+\frac{d\sigma_{rr}^{\rm stat}}{dr}\,\delta R+\delta\sigma_{rr}-2\sigma_{r\theta}^{\rm stat}\,\frac{\partial_\theta\delta R}{R}\nonumber\\
(\sigma_{\rm nt})_{r=R+\delta R} &= \sigma_{r\theta}^{\rm stat}+\frac{d\sigma_{r\theta}^{\rm stat}}{dr}\,\delta R-(\sigma_{\theta\theta}^{\rm stat}-\sigma_{rr}^{\rm stat})\frac{\partial_\theta\delta R}{R}+\delta\sigma_{r\theta}\nonumber\\
\rho_{r=R+\delta R}&=\rho^{\rm stat}+\frac{d\rho^{\rm stat}}{dr}\,\delta R+\delta\rho\nonumber\\
(v_r)_{r=R+\delta R}&=v_r^{\rm stat}+\frac{dv_r^{\rm stat}}{dr}\,\delta R+\delta v_r\, .
\end{align}
In the right-hand sides, all the quantities that depend on $r$ are evaluated at $r=R$. Note that $\sigma_{r\theta}^{\rm stat}$ is formally written here but is actually zero in the spherical symmetric case of our stationary state. Similarly, the stationary velocity $v_r^{\rm stat}$ is also zero at the outer stationary boundary $r=R$.

\section{Additional solutions of the perturbed axisymmetric mode decomposition in the limit of fast diffusion}\label{appAddModeExpression}

The components of the pressure and total stress tensor read
\begin{align}\label{eqPerturbedPressureStress}
P^{(n)} &= \left[\frac{(4n+6)\eta}{n}\,\frac{a_n}{R}+\frac{(n-6)\eta+3n\zeta}{3n}\,\kappa\,R\,c_n\right]\,\bar{r}^n\nonumber\\
\sigma^{(n)}_{rr} &= \frac{\eta}{R}\left[\left(\frac{2n^2-2n-6}{n}\,a_n-\frac{n-2}{n}\,\kappa\,R\,c_n\right)\bar{r}^n+2(n-1)\,b_n\bar{r}^{n-2}\right]\nonumber\\
\sigma^{(n)}_{r\theta} &= \frac{\eta}{R}\left[\left(-2n(n+2)\,a_n+n\,\kappa\,R\,c_n\right)\bar{r}^n-2(n^2-1)\,b_n\bar{r}^{n-2}\right]\, ,
\end{align}
where the perturbed pressure $\delta P$ and the perturbed component $\delta\sigma_{rr}$ of the stress tensor are decomposed on the basis of Legendre polynomials, and the perturbed component $\delta\sigma_{r\theta}$ of the stress tensor is decomposed on the basis of Gegenbauer polynomials\footnote{Note that the other non-trivial components of the stress tensor, $\delta\sigma_{\theta\theta}$ and $\delta\sigma_{\phi\phi}$, have expansion coefficients both on the Legendre and Gegenbauer polynomials and are not reported here.}. The three integration constants $a_n$, $b_n$, and $c_n$ are further determined by the three boundary conditions at the outer surface of the multicellular spheroid, corresponding to the continuity of the normal and tangential components of the stress tensor, as well as that of the metabolite concentration.

Solving the boundary-condition equations~\ref{eqPerturbedBoundaryFields} for the metabolite field with its expression given by equation~\ref{eqPerturbedNutrientVelocity}, we get
\begin{equation}
c_n = -\frac{1}{3}\frac{k_0R}{\kappa\,l_D^2}\,R^{(n)}\, .
\end{equation}
Doing the same for the stress tensor gives an equation for $a_n$ and $b_n$:
\begin{equation}
\begin{pmatrix}
\frac{n^2-n-3}{n} & n-1 \\
-\frac{n+2}{n+1} & -\frac{n-1}{n} 
\end{pmatrix} \cdot \begin{pmatrix}
a_n \\
b_n
\end{pmatrix} = \begin{pmatrix}
\left[1-\frac{n(n+1)}{2}\right]\frac{\gamma}{\eta\,R}-\frac{n-2}{6n}\,k_0\left(\frac{R}{l_D}\right)^2+2(\kappa\rho^{\rm ext}-k_0)\\
\frac{1}{6(n+1)}\,k_0\left(\frac{R}{l_D}\right)^2+\kappa\rho^{\rm ext}-k_0
\end{pmatrix}\,R^{(n)}\, .
\end{equation}
Using further the kinematic condition equation~\ref{eqKinematicCondition} leads to the mode growth rates given by equation~\ref{eqModeZerothOrder}.

For the modes $n=0$ and $n=1$, the expressions reported in section~\ref{sseExplicitSol} can further be expressed as functions of either $k_0$ and $\bar{R}=R/l_D$ only or $\kappa\rho^{\rm ext}$ and $\bar{R}$ only, thanks to equation~\ref{eqStatRadius}. This leads to
\begin{align}\label{eqModes0and1ZerothOrderBis}
\omega_0 &=k_0\left[\frac{\bar{R}^2}{3(\bar{R}\coth{\bar{R}}-1)}-1-\frac{1}{9}\bar{R}^2\right]=\kappa\rho^{\rm ext}\left[\frac{4}{3}+\frac{3}{\bar{R}^2}-\frac{(9+\bar{R}^2)\coth{\bar{R}}}{3\bar{R}}\right]\nonumber\\
\omega_1 &=k_0\left[\frac{\bar{R}^2}{9(\bar{R}\coth{\bar{R}}-1)}-\frac{1}{3}-\frac{1}{18}\bar{R}^2\right]=\kappa\rho^{\rm ext}\left[\frac{1}{2}+\frac{2}{\bar{R}^2}-\frac{(6+\bar{R}^2)\coth{\bar{R}}}{6\bar{R}}\right]\, .
\end{align}
With these expressions, we can easily verify that these two modes are always stable.

\section{Implicit equation for the mode $n=0$ in the generic case}\label{appImplicitEquation}

Considering the generic case of equation~\ref{eqNutrientDynamicsPerturbed}, the growth rates $\omega_n$ are given by implicit equations, which cannot be solved analytically. We report here as an example the implicit equation giving the mode $n=0$, which corresponds to the simplest one:
\begin{equation}\label{eqMode0ExactComputation}
\omega_0 =k_{\rm p}^{\rm ext}-\frac{1}{3}k_0\frac{\bar{R}^2}{\sinh{{\bar{R}_0}}}\left[\frac{\cosh{\bar{R}_0}}{\bar{R}_0}-\frac{\sinh{\bar{R}_0}}{{\bar{R}_0}^2}\right]\, ,
\end{equation}
where $\bar{R}=R/l_D$ with $l_D=\sqrt{D/\alpha}$ as before, and $\bar{R}_0=R/l_0$ with $l_0=\sqrt{D/(\alpha+\omega_0)}$.

\section{Shapes of the lowest-order instability modes}\label{appModeShapes}

\begin{figure}[t]
\begin{center}
\includegraphics[width=0.6\textwidth]{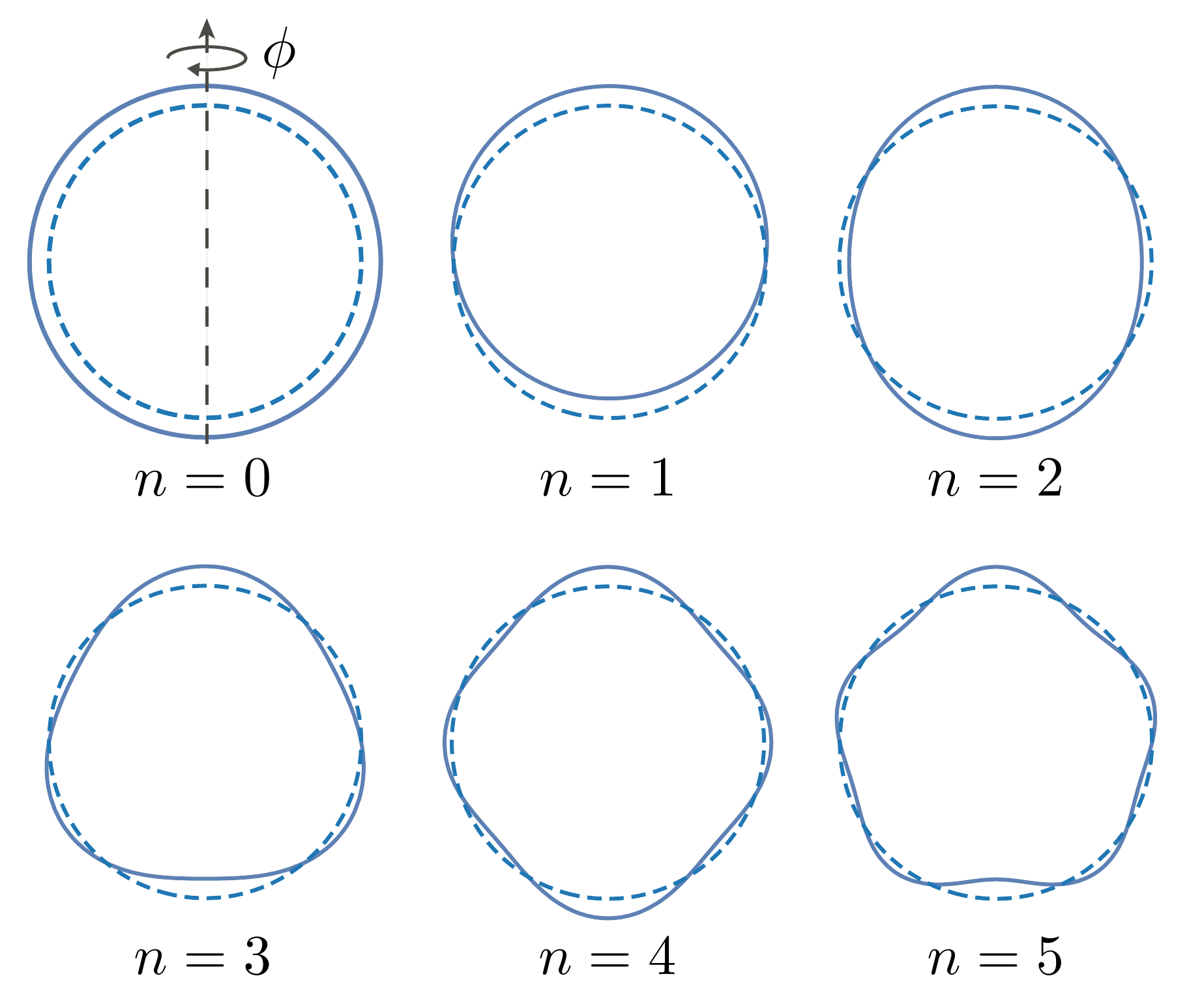}
\end{center}
\caption
{\label{figFourierModes} Schematic illustration of the lowest-order, axisymmetric, perturbative modes. The modes $n=0$ to $n=5$ are represented with an arbitrary amplitude. In each schematic representation, the unperturbed spherical shape is represented with a dashed line. In the first, $n=0$ mode, the axisymmetry is indicated as a rotational invariance in $\phi$, and all modes depicted here are invariant under this transformation.}
\end{figure}

We illustrate in figure~\ref{figFourierModes} the shapes of the lowest-order instability modes.

\section{Mode structure for intermediate values of the diffusion coefficient of metabolites}\label{appDiffusionStudy}

\begin{figure}[t]
\begin{center}
\includegraphics[width=0.75\textwidth]{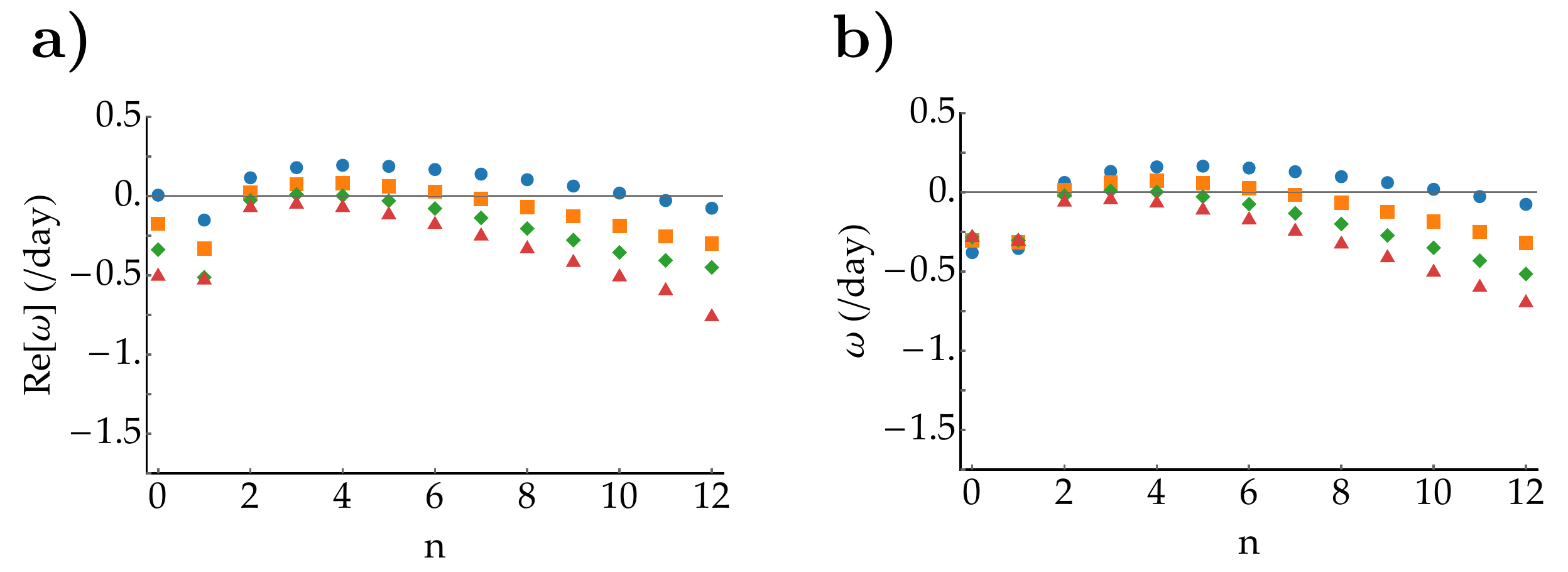}
\end{center}
\caption
{\label{figStudyDiffusion} Mode growth rates as functions of the mode number $n$ for intermediate values of the diffusion coefficient of metabolites. Parameters are the same as those used in figure~\ref{figModeResults}b,c, using the same colour code, except for the diffusion coefficient of metabolites $D$, which here takes the respective values (a) $D=10\cdot 10^{-10}$~cm$^2\cdot$s$^{-1}$ and (b) $D=20\cdot 10^{-10}$~cm$^2\cdot$s$^{-1}$, and the metabolite consumption rate $\alpha$, which is scaled accordingly to keep the same metabolite penetration lengths $l_D$ and associated steady states.}
\end{figure}

We illustrate in figure~\ref{figStudyDiffusion} the mode structure for two values of the diffusion coefficient of metabolites $D$, intermediate between those of figures~\ref{figModeResults}b,c and \ref{figModeResults}d.

\section*{References}

\providecommand{\newblock}{}

\end{document}